\documentstyle[12pt]{article}
 \advance\oddsidemargin by -1in
 \advance\evensidemargin
 by -1in
 \marginparsep
 0.4cm
 \advance\topmargin by
 -0.0in
 \makeindex

 \newcommand\la{\langle}
 \newcommand\ra{\rangle}
 \newcommand\beq{\begin{equation}}
 
 \newcommand\eeq{\end{equation}}
 \newcommand\beqn{\begin{eqnarray}}
 \newcommand\eeqn{\end{eqnarray}}
 \newcommand{\doublespace} {
 \renewcommand{\baselinestretch} {1.6}
 \large\normalsize}

\begin{document}
\vspace*{1.5cm}
 
\vspace*{4cm}
 
 \centerline{\Large \bf Baryon Stopping at HERA:}
\medskip
 \centerline{\Large \bf Evidence for Gluonic Mechanism}
 
\vspace{.5cm}
\begin{center}
{\large Boris~Kopeliovich$^{1,2}$
and Bogdan Povh$^1$}
 
\vspace{0.3cm}
 
 $^1${\sl Max-Planck Institut f\"ur Kernphysik, Postfach
103980, 69029 Heidelberg, Germany}\\
 $^2${\sl Joint Institute for Nuclear Research, Dubna,
141980 Moscow Region, Russia}
\end{center}

\vspace{1cm}
 
\begin{abstract}
Recent results from the H1 experiment \cite{h1} have confirmed
the existence of a substantial baryon asymmetry 
of the proton sea at small $x$ with magnitude
predicted in \cite{kp}. This is strong support for
the idea \cite{kz-zpc} that baryon number can be transferred
by gluons through a large rapidity interval without attenuation.
In this paper we calculate the dependence of baryon asymmetry on
associated multiplicity of produced hadrons which turns out to be
very sensitive to the underlying dynamics.
Comparison with data \cite{h1} confirms the dominance of
the gluonic mechanism of baryon number transfer and excludes
any substantial contribution of valence quark exchange.
 
\end{abstract}

\doublespace
 
\newpage
 
A sizeable baryon-antibaryon asymmetry in photon-proton interaction 
was recently observed by the H1 Collaboration for
$p/\bar p$ with small momentum in the laboratory frame produced
in $\gamma p$ collisions at HERA. The preliminary data 
presented at the Vancouver Conference \cite{h1} show that
\beq
A=2\,\frac{N_p-N_{\bar p}}{N_p+N_{\bar p}}=
(8.0\pm 1.0\pm2.5)\%\ .
\label{1}
\eeq
Here $N_p$ and $N_{\bar p}$ are the numbers of detected 
protons and antiprotons respectively.

Obviously, the observed excess of protons is a 
consequence of the presence of
the proton baryon number (BN) 
in the initial state of the reaction. Nontrivial is,
however, the very large
rapidity interval of about 8 units between the initial and final 
protons. One could expect an exponential attenuation of the BN flow
over such a long rapidity interval and a vanishing baryon asymmetry.
In contrast, a baryon
asymmetry $A$ of about $7\%$ was predicted
in \cite{kp}.  The calculations are based on the 
gluonic mechanism of BN transfer
first suggested in \cite{kz-zpc}. 
This mechanism provides 
a rapidity independent probability of BN stopping, which
is natural for gluonic exchanges. 

In topological classification \cite{rv} BN is associated with 
the string junction for a star-shaped string configuration
in the baryon. Therefore, baryon stopping is stopping of the
string junction. It is argued in \cite{rv} that at asymptotic energies
the string junction alone is stopped without valence quarks,
{\it i.e.} only gluonic fields are involved. This establishes
a correspondence to our gluonic mechanism of BN stopping.

Another mechanism of baryon stopping also suggested and calculated
within pQCD in \cite{kz-zpc} is associated with a probability
to find in the proton a
low-$x$ valence quark accompanied by the string junction.
The dependence of BN transfer on the rapidity interval $\Delta y$
is proportional to ${\rm exp}(-\Delta y/2)$ since it is related
to the well known $x$-distribution of valence
quarks dictated by Regge phenomenology \cite{roberts}. 
Evaluated in pQCD \cite{kz-zpc}
this mechanism well explains both the magnitude and
energy dependence of baryon asymmetry observed at central rapidity
in $pp$ collisions at the ISR \cite{isr1}-\cite{isr3}. 
The contribution of the
asymptotic gluonic mechanism calculated in \cite{kp}
is too small to show up at such a
small rapidity interval $\Delta y \leq 4$.

The source of baryon asymmetry can be understood in the 
parton model \cite{kp}.
In the infinite momentum frame of the proton
one can attribute a partonic interpretation to the carrier
of BN, the string junction,
since it carries a fraction of the proton momentum \cite{rv}.
In the rest frame of the proton all the partons in
the initial state of the $\gamma p$ interaction
belong to the photon. Obviously, the parton distribution of the
photon is BN symmetric. However, the interaction with 
the proton target breaks up
this symmetry due to the possibility of annihilation of anti-BN in 
the projectile parton cloud
of the photon with BN of the proton.
This leads to a nonzero BN asymmetry  in the final state.
The rapidity distribution of the produced net BN is
related to the energy behaviour of the annihilation cross section.

The annihilation cross section at very high energies was predicted
to be nearly energy independent, 
in nonperturbative \cite{gn} 
and perturbative \cite{k,kz-yaf,kz-pl}
approaches. In both cases the magnitude was predicted to be
$\sigma_{ann}(\bar pp)\approx 1 - 2\ mb$. Such a contribution was indeed
found in 
the analysis \cite{kz-pl,kz-rev} of data on multiplicity distribution
in $pp$ and $\bar pp$ collision. The corresponding cross section
$\sigma_{ann}(\bar pp)\approx 1.5\pm 0.1\ mb$
agrees well with the theoretical expectation.
This asymptotic mechanism of annihilation 
results in a rapidity independent BN transfer 
according to the above partonic picture.
Its contribution to baryon asymmetry was estimated 
in \cite{kp}. 

Experimental data for $\bar pp$ annihilation are available only
at low energy up to $12\ GeV$. In this energy range the
annihilation cross section is much larger that the asymptotic
value $1.5\ mb$ and
decreases with energy approximately as $s^{-1/2}$.
This is to be explained by the preasymptotic mechanism
corresponding to the exchange of a valence quark accompanied with the 
string junction. This assignment 
explains in a natural way the energy dependence
of annihilation,
moreover, a parameter-free evaluation of the cross section \cite{kz-ann}
is also in a good agreement with the data. 

The same preasymptotic mechanism nicely explains \cite{kz-zpc}
the energy dependence
and the absolute value for the cross section of net BN production
in the central rapidity region in $pp$ collisions at ISR.
Baryon stopping in heavy ion collisions measured at the SPS
is also well explained by this mechanism \cite{ck,vgw}
without free parameters.

This valence quark-exchange contribution to BN transfer 
in $\gamma p$ collisions was also calculated in \cite{kp}.
It turns out that the rapidity interval between the initial
and final protons in the experiment \cite{h1}
is not large enough to exclude possibility of 
this contribution. Both, the gluonic
and quark exchange mechanisms are estimated in \cite{kp}
to give about the same asymmetry at rapidity $\eta=0$
(see Fig.~5 in \cite{kp}) and
are able to explain the data within
uncertainty of calculations. 
One needs more detailed information to discriminate between
the two mechanisms.

It worth noting that a nice
topological classification  for these mechanisms
was suggested by Rossi and Veneziano \cite{rv}. 
However, their prediction for the energy dependence
has no reasonable justification and is in a 
severe contradiction with
the standard high-energy Regge phenomenology for total 
cross sections (see discussion
in review \cite{kz-rev}), therefore we discard it. 
Particularly, if the difference
between $\bar pp$ and $pp$ total cross sections is due to
annihilation, one has to eliminate the $\omega$ exchange
from the elastic $pp$ amplitude, but keep it as the major 
Reggeon term in the $Kp$ elastic scattering. Another problem
arises with relation between $\omega$ and $\rho$ exchanges, since
the latter is predicted by quark models to be much smaller than 
the former in agreement with results of the standard Regge 
phenomenology. According to Eylon and Harari \cite{eh}
annihilation is related via unitarity to the Pomeron
(at least a substantial part of it, see \cite{kz-ann})
and does not contribute to the $\bar pp$ and $pp$ total cross section
difference.

\begin{figure}[tbh] \includegraphics{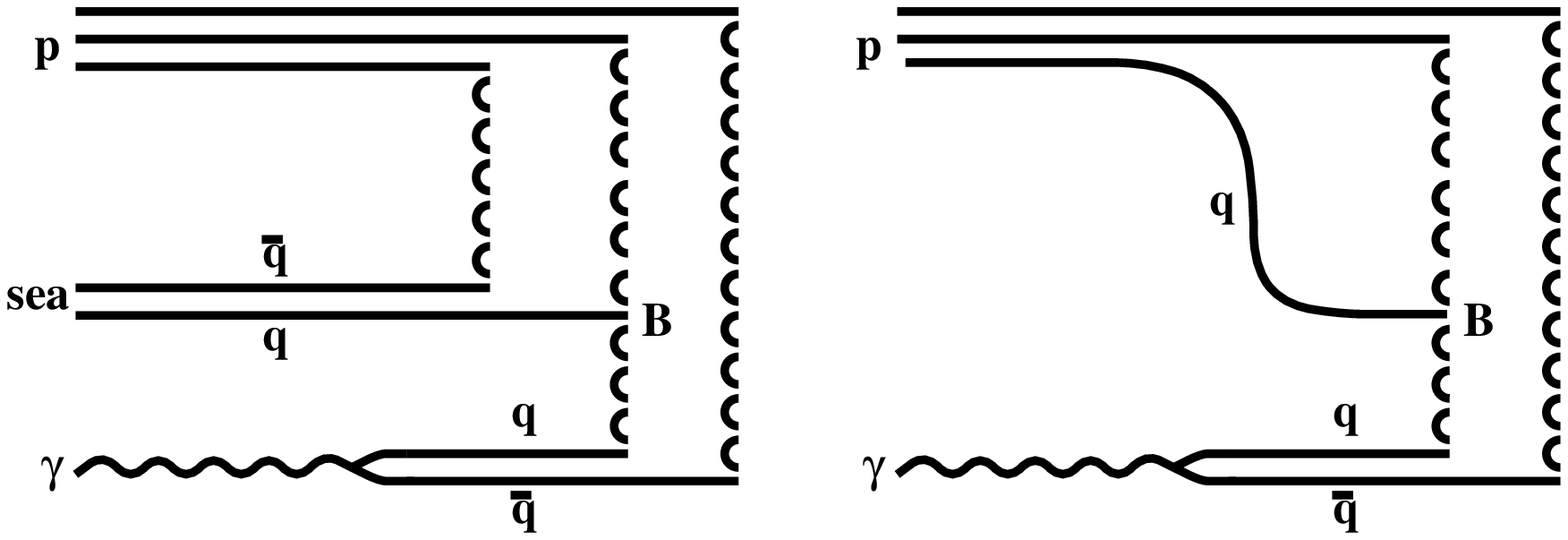}
\begin{center} 
\vspace{4cm} 
\parbox{13cm} 
{\caption[Delta]
{\sl Multiparticle production corresponding to the 
gluonic mechanism of BN transfer in $\gamma-p$ interaction.
The BN is produced with the rapidity of the sea quark (left)\\
The same as on the left, but for the valence quark mechanism.
BN has the rapidity of the valence quark (right).}
\label{fig1}} 
\end{center} 
\end{figure}

An important signature of the asymptotic gluonic mechanism is
a higher mean multiplicity of produced particles. This
is due to three sheet topology of the final state according to
classification in \cite{rv}, which is illustrated 
in Fig.~\ref{fig1} (left).
The gluon is replaced by a sea $\bar qq$ pair.
The valence quark
exchange mechanism also shown in Fig.~\ref{fig1} exhibits a 
two-sheet (two-string) 
topology, {\it i.e.} the same multiplicity as in the Pomeron.
It is easy to see in Fig.~\ref{fig1} 
that the mean multiplicity of produced particles
in the rapidity interval where the baryon asymmetry is measured
is 5/4 times larger for the gluonic mechanism (left) compared
to the quark exchange mechanism (right). 
This fact makes baryon asymmetry dependent on the 
multiplicity of the produced hadrons. 

\begin{figure}[tbh] 
\includegraphics{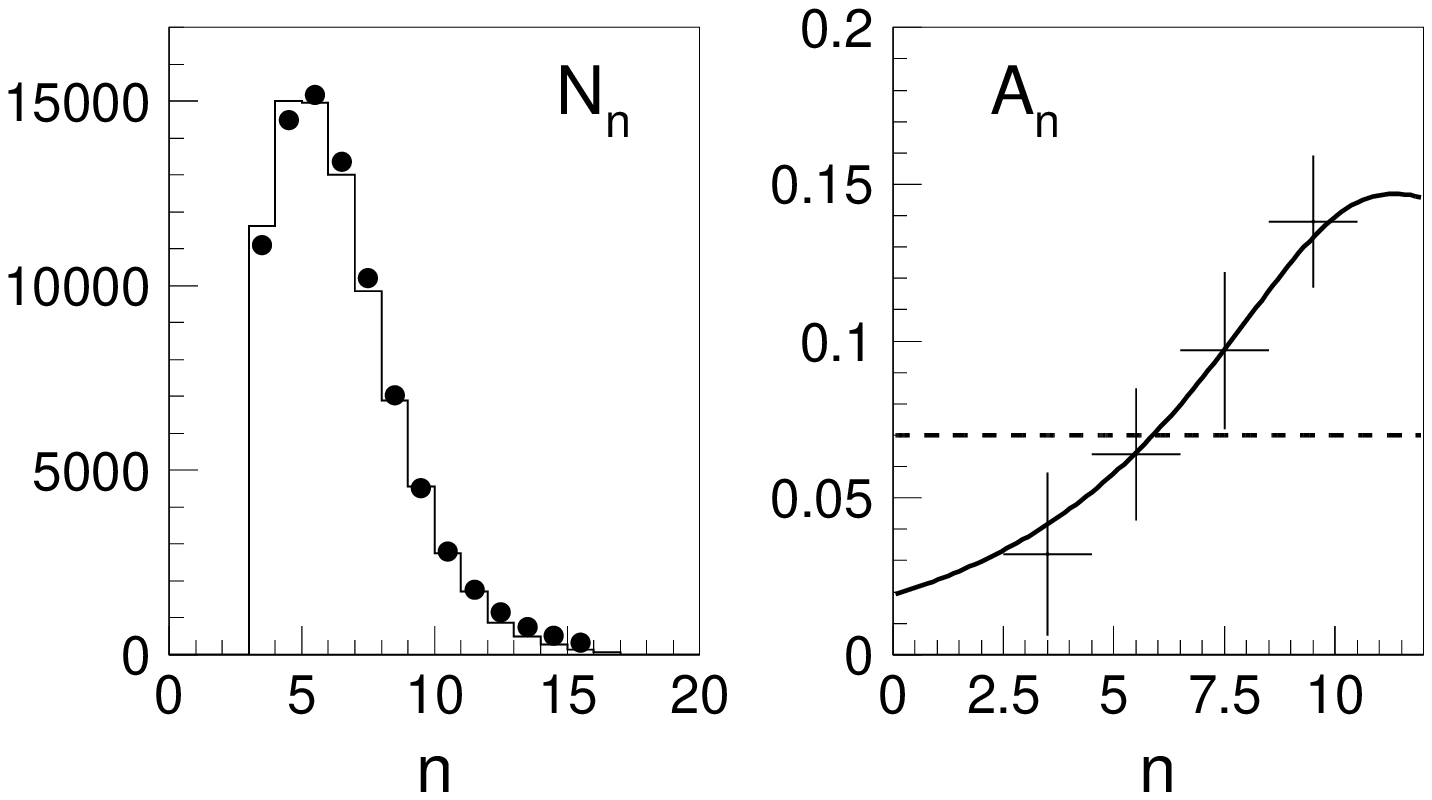}
\begin{center}
\vspace{8.5cm}
\parbox{13cm}
{\caption[Delta]
{\sl Multiplicity distribution of charged hadrons produced 
in photon-proton interaction
as measured in \cite{h1}. The histogram represents the data,
the black points are the result of our calculation (left)\\
Baryon asymmetry as function of multiplicity of charged hadrons.
The crosses are the results of measurements in \cite{h1}.
The solid and dashed curves show our predictions for the
gluonic and quark exchange mechanisms respectively (right).
See the text for details}
\label{fig2}}
\end{center}
\end{figure}

First of all, we should describe the multiplicity distribution
measured in \cite{h1}. We use the standard AGK cutting rules
\cite{agk} which relate via unitarity 
inelastic processes with multi-pomeron exchanges in the elastic 
amplitude. Keeping only the double-Pomeron corrections the
multiplicity distribution normalized to unity reads,
\beq
N_n = 
(1-4\delta)\frac{\la n\ra^n}{n\,!}\,e^{- \la n\ra} +
4\delta \frac{(2\la n\ra)^n}{n\,!}\,e^{- 2\la n\ra}\ .
\label{2}
\eeq
Here $\la n\ra$ is  the mean
number of produced particles corresponding to one cut pomeron;
the parameter $\delta$ is the weight of the double-pomeron contribution 
to the total cross section.
We use them as free parameters to adjust to the data by eye.
The result with $\la n\ra=5.2$ and $\delta=0.02$ 
is shown in Fig.~\ref{fig2} (left) by full circles. 
It is remarkable that the value of $\delta$ is 
about an order of magnitude smaller than what 
follows from eikonal model, which is
reasonably good for hadronic collisions. We interpret it
as suppression of gluons and sea quarks, 
which are responsible for the 
multi-pomeron terms, in a photon compared to a hadron. 
This would also naturally explain a substantially steeper growth with
energy of the photon-photon total cross section measured
by the L3 collaboration \cite{l3}.
A further development of this issue
goes too far beyond the scope of present paper.
Whatever the reason of smallness of $\delta$ is, 
it has no effect on our results
for multiplicity dependence of baryon asymmetry.

Now we are in position to predict  the variation of 
the baryon asymmetry 
(\ref{1}) with associated multiplicity.
Most of the $p$ and $\bar p$ contributing to the denominator
of (\ref{1}) are due to production of sea baryon-antibaryon
pairs, number of which is proportional to the multiplicity 
of hadrons. Therefore, the shape of $n$-dependence of
$N_p+N_{\bar p}$ in (\ref{1}) is given by (\ref{2}).
The same is true for the numerator in (\ref{1}), except
the mean multiplicity associated with net BN production
according to Fig.~\ref{fig1} is larger for the gluonic mechanism.
Using (\ref{2}) we can represent the asymmetry (\ref{1})
as function of $n$,
\beq
A_n = A\,\frac{(1-4\delta)\,K^n\,e^{-(K-1)\la n\ra}+
4\delta\,(K+1)^n\,e^{-K\la n\ra}}
{1+4\delta(2^n\,e^{-\la n\ra} - 1)}
\label{3}
\eeq
According to the previous discussion we
assume that the mean associated multiplicity for net BN production
(see Fig.~\ref{fig1}) is $K$ times larger than the mean multiplicity 
in BN symmetric events,
where $K=5/4$ or $K=1$ for gluonic and quarks exchange mechanisms
respectively.
The results for $A_n$ are depicted
in Fig.~\ref{fig2} (right) by solid ($K=5/4$) and dashed ($K=1$)
lines. 
We use $A=0.07$ as it is predicted in \cite{kp}.
The data \cite{h1} shown by crosses agree well with the assumption
that the baryon asymmetry is dominated by the contribution of the 
gluonic mechanism, but reject any sizeable contribution
of the preasymptotic quark exchange mechanism which 
leads to a constant $A_n$. 

Note, that the systematical uncertainty
not included in the error bars in Fig.~\ref{fig2} can affect
only the overall normalization \cite{h1}, but does not
change the shape of $n$-dependence. The kinematical restrictions
imposed by the experimental set up do not affect our calculations,
which depend only on parameters $\la n\ra$ and $\delta$ measured in
the same experiment.

Summarizing, an exciting possibility that baryon number can be 
transferred by gluons through a large rapidity interval 
without attenuation \cite{kz-zpc}
is strongly supported by 
measurements \cite{h1} of baryon asymmetry in $\gamma p$ collisions.
Although the magnitude of the effect 
agrees with the prediction made in \cite{kp},
this fact alone cannot exclude a large contribution of the preasymtotic
valence quark exchange mechanism at this rapidity interval
$\Delta y \approx 7\,-\,8$. We have found that the dependence of 
baryon asymmetry on associated particle multiplicity is extremely
sensitive to the underlying mechanism. Comparison with
corresponding data from \cite{h1} strongly supports dominance
of the gluonic mechanism and excludes a large contribution of
BN transfer by valence quarks.

\medskip
 
{\bf Acknowledgements}: We are grateful to Andrei Rostovtsev for 
useful discussions.

\end{document}